\documentstyle[12pt]{article}
\title{\huge The Effect of off-ramp on the one-dimensional cellular automaton traffic flow with open boundaries } 
  
\author{\bf Hamid Ez-Zahraouy$^*$, Zoubir Benrihane, Abdelilah Benyoussef  
\\
 Laboratoire de Magn\'{e}tisme et de la Physique
 des Hautes Energies
\\
Universit\'{e} Mohammed V, Facult\'{e} des Sciences, Avenue Ibn Batouta,  B.P. 1014
\\Rabat, Morocco
}
\date{ }

\begin{document}
\maketitle
\abstract{The effect of the off-ramp (way out),  on the traffic flow phase transition is investigated using numerical simulations in the one-dimensional cellular automaton traffic flow model with open boundaries using parallel dynamics. When the off-ramp is located  between two critical positions $ i_{c1}$ and $ i_{c2}$  the current increases with the extracting rate $\beta_{0}$, for $\beta_{0}<\beta_{0c1}$, and exhibits a plateau (constant current) for $\beta_{0c1}<\beta_{0}<\beta_{0c2}$ and decreases with $\beta_{0}$ for $\beta_{0}>\beta_{0c2}$. However, the density undergoes two successive first order transitions; from high density to plateau current phase at $\beta_{0}=\beta_{0c1}$ ; and from average density to the low one at $\beta_{0}=\beta_{0c2}$. In the case of two off-ramps located respectively at  $ i_{1}$ and $ i_{2}$, these transitions occur only when $i_{2}-i_{1}$. Phase diagrams in the $(\alpha,\beta_{0}), (\beta,\beta_{0})$  and  $(i_{1},\beta_{0})$ planes are established. It is found that the transitions between free traffic (FT), Congested traffic(CT) and plateau current (PC) phases are of first order. The first order line transition in $(i_{1},\beta_{0})$phase diagram terminates by an end point above which the transition disappears.}\\\\
Pacs number : 89.40.-a, 05.50.+q , 64.60.Cn , 82.20.wt\\
Keywords :\rm\ Traffic flow, Open boundaries, Numerical simulation, off-ramp, Phase diagrams.\\
\------------------------------------------\\
$^*$corresponding author E-mail address: ezahamid@fsr.ac.ma

\newpage

\section{\protect\bigskip Introduction}

In the past few years, traffic problems have attracted the interest of a community of physicists [1-3]. Traffic flow, a kind  of many-body systems of strongly interacting vehicles, shows various complex behaviors. Numerous empirical data of the highway traffic have been obtained, which demonstrate the existence of qualitatively distinct dynamic states [4-7]. In particular, three distinct dynamic phases are observed on highways: the free traffic flow, the traffic jam, and the synchronized traffic flow. It has been found out experimentally that the complexity in traffic flow is linked to diverse space-time transitions between the three basically different kinds of traffic [4].

To understand the behavior of traffic flow, various traffic models have been proposed and studied, including car-following models, gas-kinetic models, hydrodynamic models, and cellular automaton (CA) models [8-17], these later models have been applied to the various traffic problems for the automobile on the highway and the streets. First proposed by Nagel and Schreckenberg [9] and subsequently studied by other authors using a variety of techniques. In the (CA) models, cares are treated as distinguishable particles and roads are expressed by discrete lattices. The whole system evolves in discrete time steps. All cars can move ahead simultaneously at each time step. These treatments for the behavior of cars make analyses of the complex systems with many degrees of freedom. With the help of these models, free flow and jams are well understood. On the other hand, the nature of synchronized traffic flow remains unclear despite various efforts [4,5,16,18].

Recent experimental investigation shows that in the majority of cases, synchronized traffic is observed localized near bottlenecks and thus it is believed that bottlenecks are important for the formation of synchronized traffic. The bottlenecks include on-ramps, off-ramps, lane closings, up-hill gradients, narrow road sections, etc. Among the various types of bottlenecks, the on-ramp is of particular interest to researchers and has been widely studied [16,18-21]. Popkov et al. [22] compared the experimental data from an on-ramp with simulations of a cellular automaton model. It is concluded that the dynamics due to the ramp can be described by a bottleneck. Diedrich et al [19] and Campari and Levi [20] simulated the on-ramp using the CA models. Traffic phenomena such as synchronized flow, the lane inversion, and phase separation were reproduced.

The effects of the way(s) out (off-ramps) have been largely studied (empirically and numerically) in several recoveries. Our principal aim in these works is to answer to equation "where we can place the way(s) out and with which rate of absorption in order to have a better fluidity?» Using numerical simulations, we study the interaction between the traffic flow on the main road and way(s) out, i.e. the effect of the extracting rate $\beta_{0}$, in both one way out (one off-ramp) and two ways out (two off-ramps) in a road, on the dynamical jamming transition, density , current, and phase diagrams in the cellular automaton of one-dimensional traffic flow model with open boundaries using parallel update. It is found that the current exhibits a plateau for special positions of the way out. In spite of the simplicity of our model, it could have replicated the different behavior as well as phases fundamental of the traffic flow. CA models are conceptually simpler, and can be easily implemented on computers for numerical investigations.

The paper is organized as follows; in the following section we define the model, the section 3 is reserved for results and discussions, the conclusion is given in section 4.

\section{Model}

Motions of cars and interactions between cars are the microscopic processes in the traffic flow. One of the approaches to microscopic traffic processes is based on cellular automata. A CA model treats the motion of cars as hopping processes on one-dimensional lattices. The Wolfram's rule 184 CA is the simplest choice [23]. Nagel and Schreckenberg introduced their CA model (NS model) by extending the 184 CA to consider the high velocity and the stochastic processes [9]. They showed that start-stop waves appear in the congested traffic region as observed in real freeway traffic. In this paper, we use the exclusion asymmetric models to serve as the basis model for the implementation of way(s) out.

We consider a one-dimensional lattice of length L. Each site either occupied by one particle or is empty. A configuration of the system is characterized by binary variable ${\tau_{i}}$ where $\tau_{i}= 0 $ ($\tau_{i}= 1 $) if site i is empty (full). To discuss the implementations of way(s) out (off-ramp(s)), we assume that the main road is single lane and way(s) out connects the main road only on one lattice $i_{1}$ in the case of one way out, and two lattices $i_{1}$ and $i_{2}$ in the case of two ways out. During a time step $\Delta t$, each particle in the system has a probability p of jumping to the empty adjacent site on its right (and does not move otherwise), a vehicle can enter without constraint, with a probability $\alpha$, in the first site being to the left side of the road if this site is empty. While, a vehicle being on the right in the last site can leave the road with a probability $\beta$ and removed on the way(s) out (off-ramp(s)) with a probability $\beta_{0}$. In our simulations the rule described above is updated in parallel ($\Delta t = 1.$)[17], i.e. during one  update step the new particle positions do not influence the rest and only previous positions have to be taken into account. In order to compute the average of any parameter u $(<u>)$, the values of $u(t)$ obtained from $5\times 10^{4}$ to $10^{5}$ time steps are averaged. Starting the simulations from random configurations, the system reaches a stationary state after a sufficiently large number of time steps. In all our simulation we averaged over 50-100 initial configurations. The advantage of parallel update, with respect to sub-lattice or sequential update is that all sites are equivalent, which should be the case of a realistic model with translational invariance.\\
Thus, if the system has the configuration $\tau_{1}(t)$,$\tau_{2}(t)$,...,$\tau_{L}(t)$ at time t it will change at time $ t+ \Delta t $ to the following, in the case of one off-ramp: \\
For $1<i<L$ and $i\neq i_{1}$, 
$\tau_{i}(t+ \Delta t$) = 1 with probability $p_{i}=\tau_{i}(t)+[\tau_{i-1}(t)(1-\tau_{i}(t))-\tau_{i}(t)(1-\tau_{i+1}(t)]\Delta t$ and 
$\tau_{i}(t+ \Delta t$) = 0 with probability $1-p_{i}$.\\
Where $\Delta t$=1 for parallel update.\\ 
For $i=1$,
$\tau_{1}(t+ \Delta t$) = 1 with probability $p_{1}=\tau_{1}(t)+[\alpha(1-\tau_{i}(t))-\tau_{i}(t)(1-\tau_{2}(t)]\Delta t$ and $\tau_{1}(t+ \Delta t$) = 0 with probability $1-p_{1}$.\\
For $i=i_{1}$ and/or $i=i_{2}$, 
$\tau_{i}(t+ \Delta t$) = 1 with probability 
$p_{i}=\tau_{i}(t)+[\tau_{i-1}(t)(1-\tau_{i}(t))-\beta_{0}\tau_{i}(t)]\Delta t$ and 
$\tau_{i}(t+ \Delta t$) = 0 with probability $1-p_{i}$.\\
Where $i_{1}$ and $\beta_{0}$ denote the position(s) of off-ramp(s)and the extracting rate, respectively. \\
For $i=L$, 
$\tau_{L}(t+ \Delta t$) = 1 with probability 
$p_{L}=\tau_{L}(t)+[\tau_{L-1}(t)(1-\tau_{L}(t))-\beta\tau_{L}(t)]\Delta t$ and 
$\tau_{L}(t+ \Delta t$) = 0 with probability $1-p_{L}$.

\section{Results and Discussion}

In the case of one way out ( one off-ramp) and in order to show the combinative effect of $\alpha$, $i_{1}$ and $\beta_{0}$ on the traffic behavior, the variation of the average density as a function of $\beta_{0}$ for several values of $\alpha$ is represented in Fig.1a, we distinguish three regions when $\alpha$ is varied, a region in which we have a low value of density and does not depend upon $\beta_{0}$ when $\alpha<\beta$, and when increasing $\alpha$, high density and plateau current (PC) (the current undergoes a plateau as a function of $\beta_{0}$ ) take place. We note that the MC phase  appears between two critical values of $\beta_{0}$ ($\beta_{0c1}$ and $\beta_{0c2}$), and disappears completely when $\alpha$ becomes greater than a critical value contrary to the case of the one dimensional asymmetric exclusion model with open boundaries with parallel dynamics ($\Delta t = 1.$)[17] in which the MC phases disappears completely. However, the system exhibits three phases namely free traffic phase (FT), congested traffic (CT) and plateau current phase (PC); the transitions between these phases are first order because of the discontinuity of the global mean density at the transition. The variation of the global mean density as a function of the extracting rate $\beta_{0}$ for $\alpha=0.4$, $\beta=0.1$, and for various positions of way out is given in Fig.1b. It is clear that the system exhibits three different phases, namely, FT, CT and MC. Indeed, there exist a critical position of the way out $ i_{c1}$ below which the density is constant for $\beta_{0}<\beta_{0c2}$ and decreases discontinuously when increasing $\beta_{0}$ for $\beta_{0}>\beta_{0c2}$. When the way out is located at a position greater than $ i_{c2}$, the density decreases with $\beta_{0}$ for $\beta_{0}<\beta_{0c1}$ and becomes constant for any value of $\beta_{0}$ greater than $\beta_{0c1}$. While, when the way out is located at any position between $ i_{c1}$ and $ i_{c2}$ ($ i_{c1}< i_{c2}$), the density decreases, for $\beta_{0}<\beta_{0c1}$, with $\beta_{0}$ and becomes constant for $\beta_{0c1}<\beta_{0}<\beta_{0c2}$ and decreases with $\beta_{0}$ for $\beta_{0}>\beta_{0c2}$. In the later case the density undergoes two successive first order transitions; from high density to intermediate one (plateau current phase) at $\beta_{0}=\beta_{0c1}$ and from intermediate density to the low one at $\beta_{0}=\beta_{0c2}$. In order to show the flow behavior under the effect of $\alpha$ and the positions of off-ramp, Fig.2a present the variation of the average flow as function of $\beta_{0}$ for several values of $\alpha$ with $\beta = 0.1$ and $i_{1} = 200$, we show that the traffic flow increases with $\alpha$, passes through a maximum at $\beta_{0}=\beta_{max}$ for high values of $\alpha$ and decreases for any value of $\beta_{0}$ greater than $\beta_{max}$. Indeed, for a certain values of $\alpha$ the current increase and  becomes constant between two values of $\beta_{0}$ and decreases for $\beta_{0}$ less than a critical values. Fig.2b give for fixed values of $\alpha$ and $\beta$, an indication of how much the average current change as $\i_{1}$ is modified. When $i_{1}<i_{c1}$ The current remain constant for $\beta_{0}<\beta_{0c2}$ and decrease when increasing $\beta_{0}$. Moreover for $\beta_{0}<\beta_{0c1}$ the current increase and becomes constant for $i_{1}>i_{c2}$. While, when the way out is located at any position between $i_{c1}$ and $ i_{c2}$, the current undergoes three successive regions when $\beta_{0}$ is varied, a region in which the current increase for $\beta_{0}<\beta_{0c1}$ and remain constant between $\beta_{0c1}$ and $\beta_{0c2}$ and decreases for $\beta_{0}>\beta_{0c2}$. The system studied exhibits three phases: FT, CT and PC. The nature of the transition between these phases depends upon the values of $\alpha$, $\beta$, $\beta_{0}$ and $\i_{1}$. Collecting the results illustrated in Fig.1 and Fig.2 we obtain the phase diagrams as shown in Fig.3. The low and high density phases are separated by a first order transition line. Each of these phases undergoes discontinuous transition to the plateau current phase as show in Fig.3a. To illustrate the effect of $\beta$ for fixed value of $\alpha$ we give the phase diagram in the ($\beta,\beta_{0}$) plan in Fig.3b, we show that exist three phase similar in precedent phases and the transition between them is also the first order type. In the same way, one established another phase diagram but this time in the space $(\i_{1},\beta_{0})$ as show in Fig.3c. In the case of one way out, and for a fixed value of $\alpha$ and $\beta$ the system exhibits three phases, namely, congested traffic, free traffic and plateau current phases. The first order transition between the jamming phase and the moving one occurs at $\beta_{0}<0.72$ and the way out position lower than $i_{c1}=100$. The same nature of transition occurs at $\beta_{0}<0.37$ and the way out position greater than $i_{c2}=350$. While, when the way out is located at any position between $i_{c1}$ and $i_{c2}$, the density undergoes two successive first order transitions; from high density to intermediate one (plateau current phase) at $\beta_{0}=0.37$ and from intermediate density to the low one at $\beta_{0}=0.72$. For a larger values of $\alpha$ the plateau current disappears and the system exhibits a first order transition between high and low densities phases as Fig.3d and Fig.3e. In the case of Fig.3e the line transition terminates by an end critical point above which there is no distinction between the phases. In order to show the space-time evolution and the different region maps inside the chain. Moreover, the spatial vehicle density in the case of one off-ramp placed in the middle site of the chain depends on the values of $\alpha$, $\beta$ and $\beta_{0}$. However for $\alpha = 0.4$, $\beta = 0.1$ the system is in high (low) density phase for low (high) values of $\beta_{0}$; while for intermediate value of $\beta_{0}=0.6$ the coexistence of the free flow and jam takes place, as it is shown in the space-time evolution (Fig.4).

In order to investigate the contribution that the extracting rate, $\beta_{0}$, affects upon the density inside the chain. Fig.5 gives for fixed values of $\alpha$ and $\beta$, an indication of how much the average occupation of each position i $( \rho(i)= <\tau_{i}>)$ changes as $\beta_{0}$ is modified. The observed shifts may be interpreted as follows. For low values of $\beta_{0}$($\beta_{0} = 0.2$), the average occupation is constant in the right side in the chain and will undergo a jump spontaneously in the position of the off-ramp and remain constant in the left one, for intermediate values of $\beta_{0}$($\beta_{0} = 0.5$), the value of average occupation remains constant for any positions less than the position of off-ramp and increase smoothly in the high value of average occupation then it becomes constant, for very large value of $\beta_{0}$($\beta_{0} = 0.8$) $ \rho(i)$ remains unchanged for low position less than the position of off-ramp and decrease discontinuously in low value of occupation. We observe the same behavior in the case of two ways out with two jumps to the level of the position of the ways out.
      
In the case of two ways out located symmetrically with regard to the center respectively at the positions $i_{1}$ and $i_{2}$, when the distance between two ways out is less than a critical value $d_{c}$ the traffic flow increases with $\beta_{0}$, and becomes constant at a critical value $\beta_{0c'}$ and decreases at an other critical value $\beta_{0c"}$; increasing the distance between two ways out, the current increases with $\beta_{0}$, passes through a maximum at $\beta_{0}=\beta_{max}$ and decreases for any value of $\beta_{0}$ greater than $\beta_{max}$. However, the two successive transitions occur only when the distance $i_{2}-i_{1}$ separating the two ways is smaller than a critical distance $d_{c}$, and the system exhibits an inversion point at $\beta_{0}=\beta_{0i} = 0.4$ for different values of distance between two ways out, as show in Fig.6a. Indeed, with increasing the distance  $i_{2}-i_{1}$ the current decreases for $\beta_{0}>\beta_{0i}$ , while it increases for $\beta_{0}<\beta_{0i}$. The inversion point disappears when the position of the first way out is located at the middle site; hence, the current increases with $\beta_{0}$ and/or the distance $i_{2}-i_{1}$, passes through a maximum at $\beta_{0}=\beta_{max}$ and decreases for any value of $\beta_{0}$ greater than $\beta_{max}$ (Fig.6b). The values of $\beta_{0c1}$, $\beta_{0c2}$, $i_{c1}$, $ i_{c2}$ and $d_{c}$ depend on the injecting rate $\alpha$, the extracting rate $\beta$ and the position(s) of the way(s) out in the road. Moreover $i_{c1}$ and $ i_{c2}$, depend on the size of the system. 

\section{Conclusion}

Using numerical simulation method, we have studied the effects of the extracting rate $\beta_{0}$, in both one way out (one off-ramp) and two ways out (two off-ramps) in a road, on the traffic flow phase transition, in the one-dimensional cellular automaton traffic flow model with open boundary conditions. We have shown that the behavior of density and current depends strongly on the value of $\beta_{0}$, the position of the way out from the entering and the distance between the ways out. In the case of one way out, and for a fixed value of $\alpha$ and $\beta$ the system exhibits three phases, namely, moving phase, jamming phase and plateau current phase. The PC phase occurs only at special positions of the off-ramp(s) with special values of the extracting rate. The transitions between different phases are first order. Furthermore we have shown that the system exhibits an end point in the (off-ramp position, extracting rate) plane in the case of large value of the injecting rate. In the case of two off-ramps $i_{1}$ and $i_{2}$, the MC appears only when the distance between them is smaller than a critical value which depends on the absorbing, injecting and extracting rates.    
\\
\par\bf  ACKNOWLEDGMENT \\
\par \rm  This work was financially supported by the Protars II n° P11/02 \\

\newpage \textit{Figure captions :}

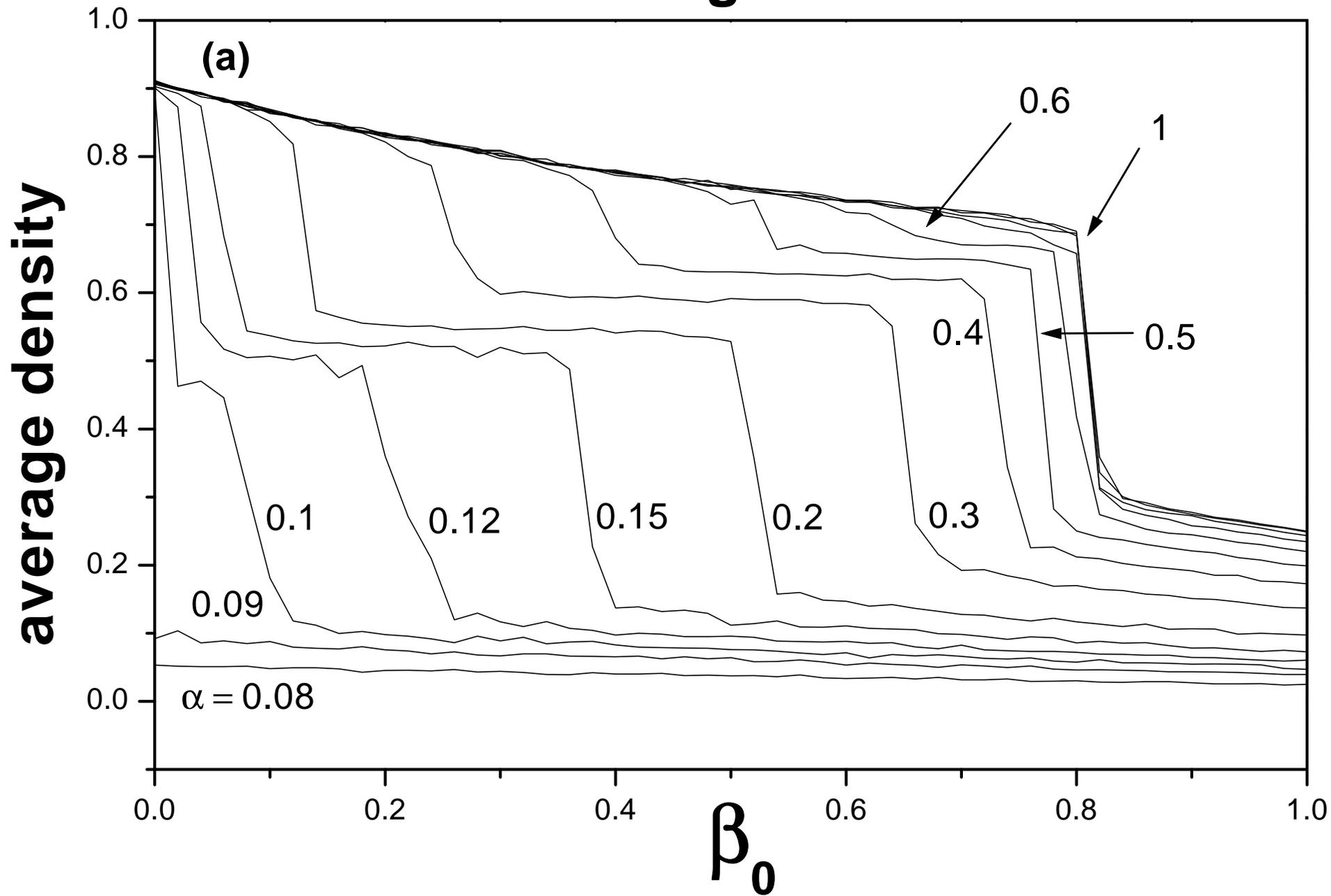
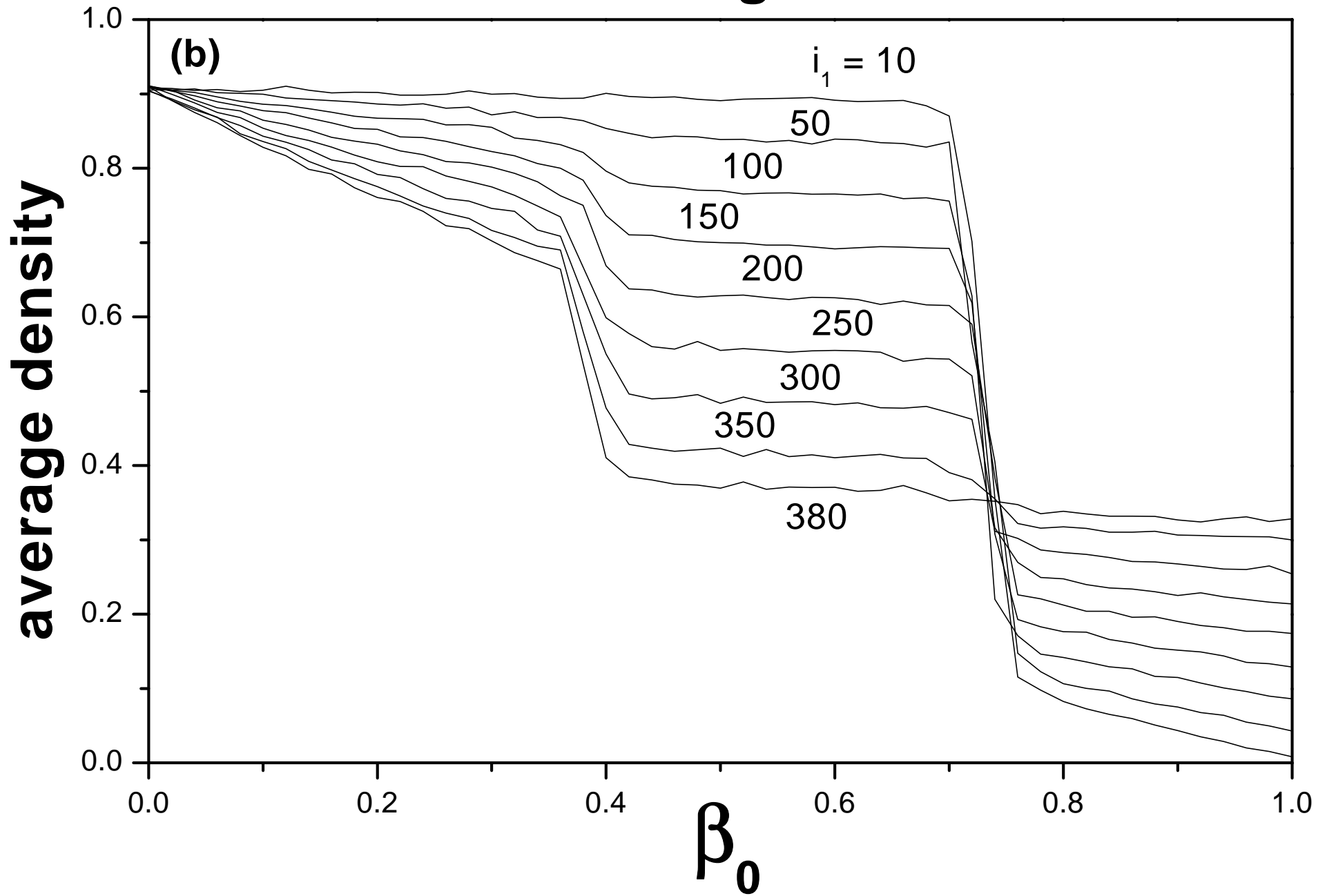
Fig.1: The variation of average density $\rho$ as a function of $\beta_{0}$, (a) $i_{1} = 200$ and  for different values of $\alpha$, (b) $\alpha=0.4$ and  for different values of $i_{1} $ , with $\beta = 0.1$ and $L = 400 $

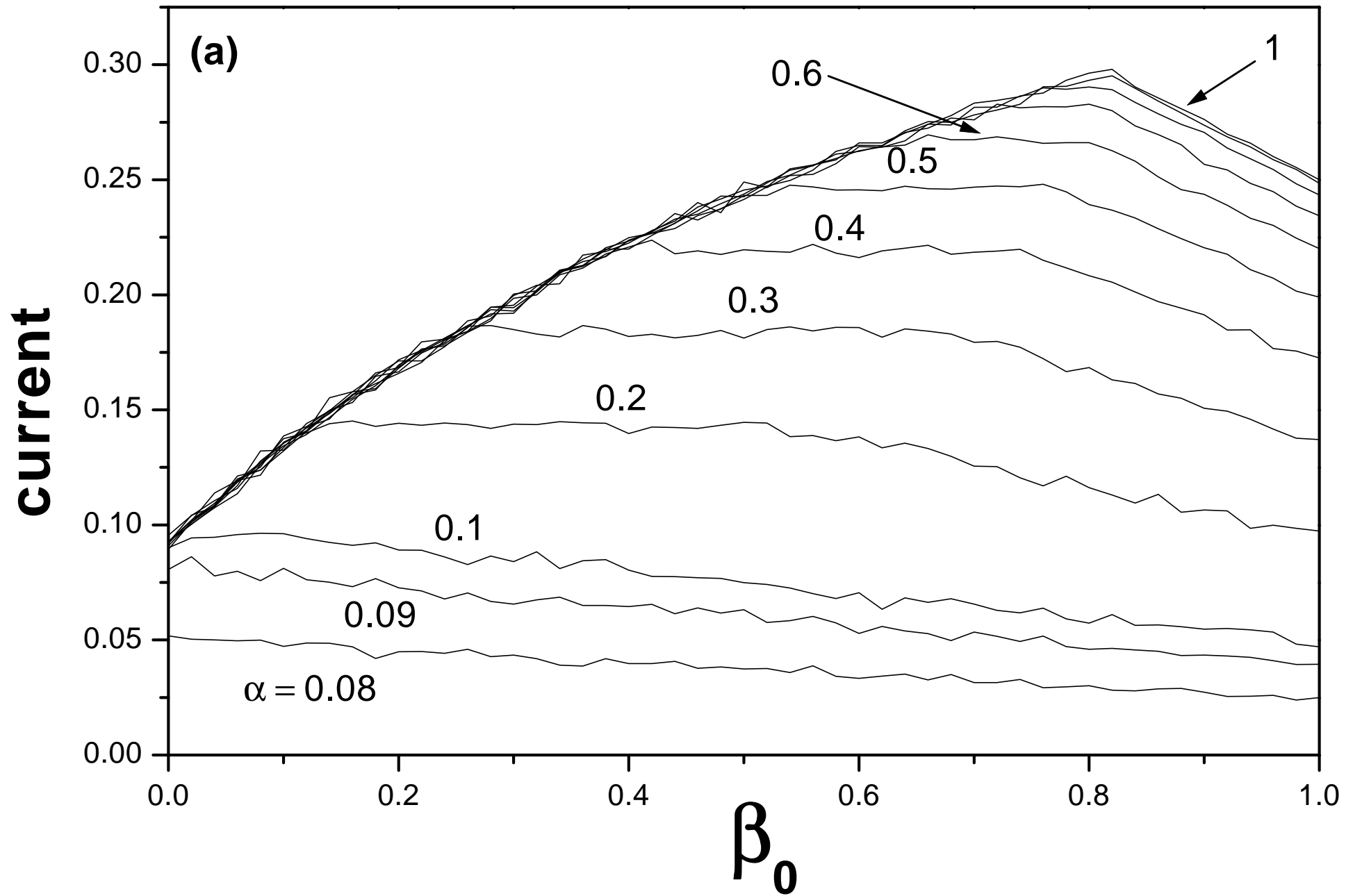
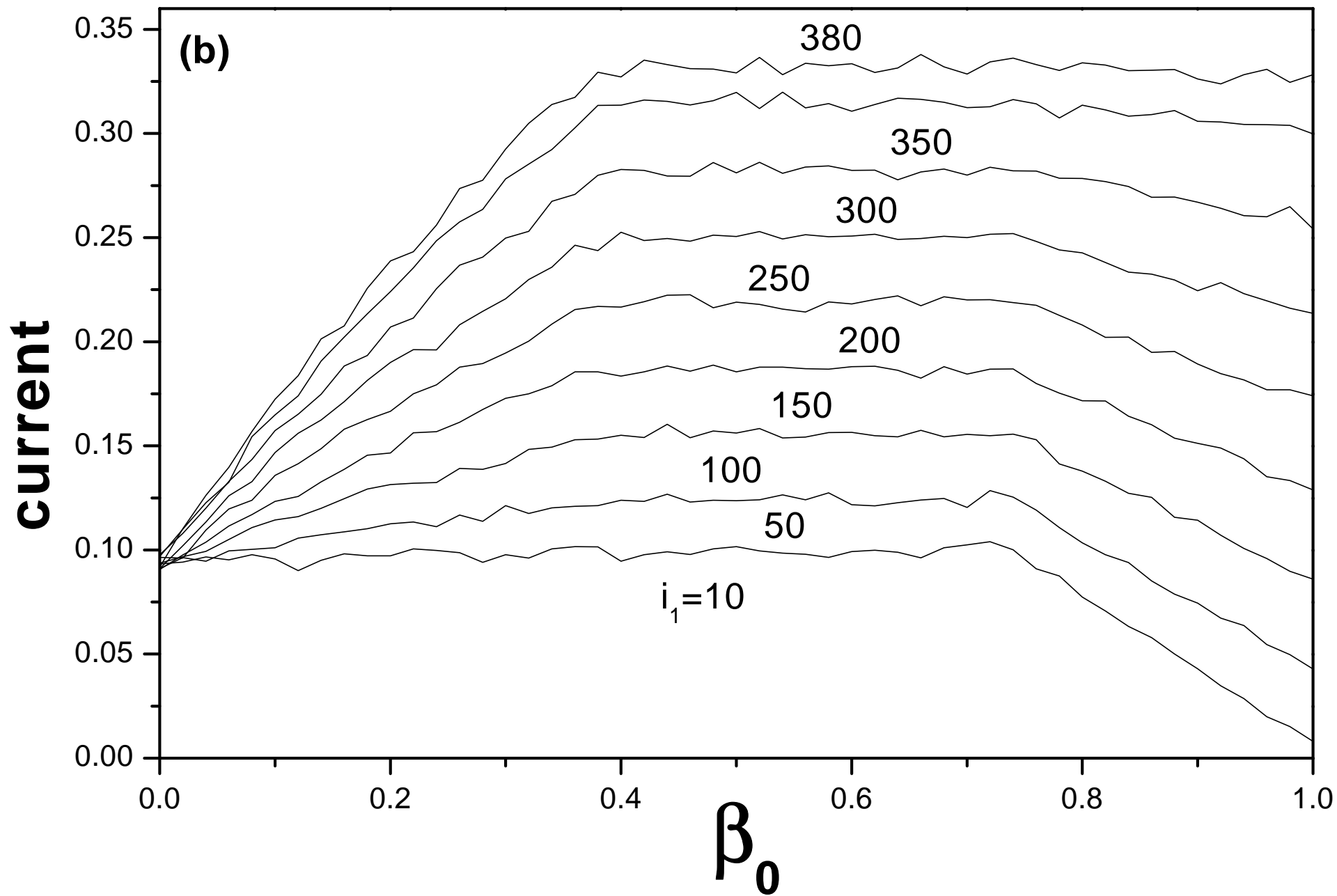
Fig.2: The variation of current J as a function of $\beta_{0}$, (a) $i_{1} = 200$ and  for different values of $\alpha$, (b) $\alpha=0.4$ and  for different values of  $i_{1}$, with $\beta = 0.1$ and $L = 400 $

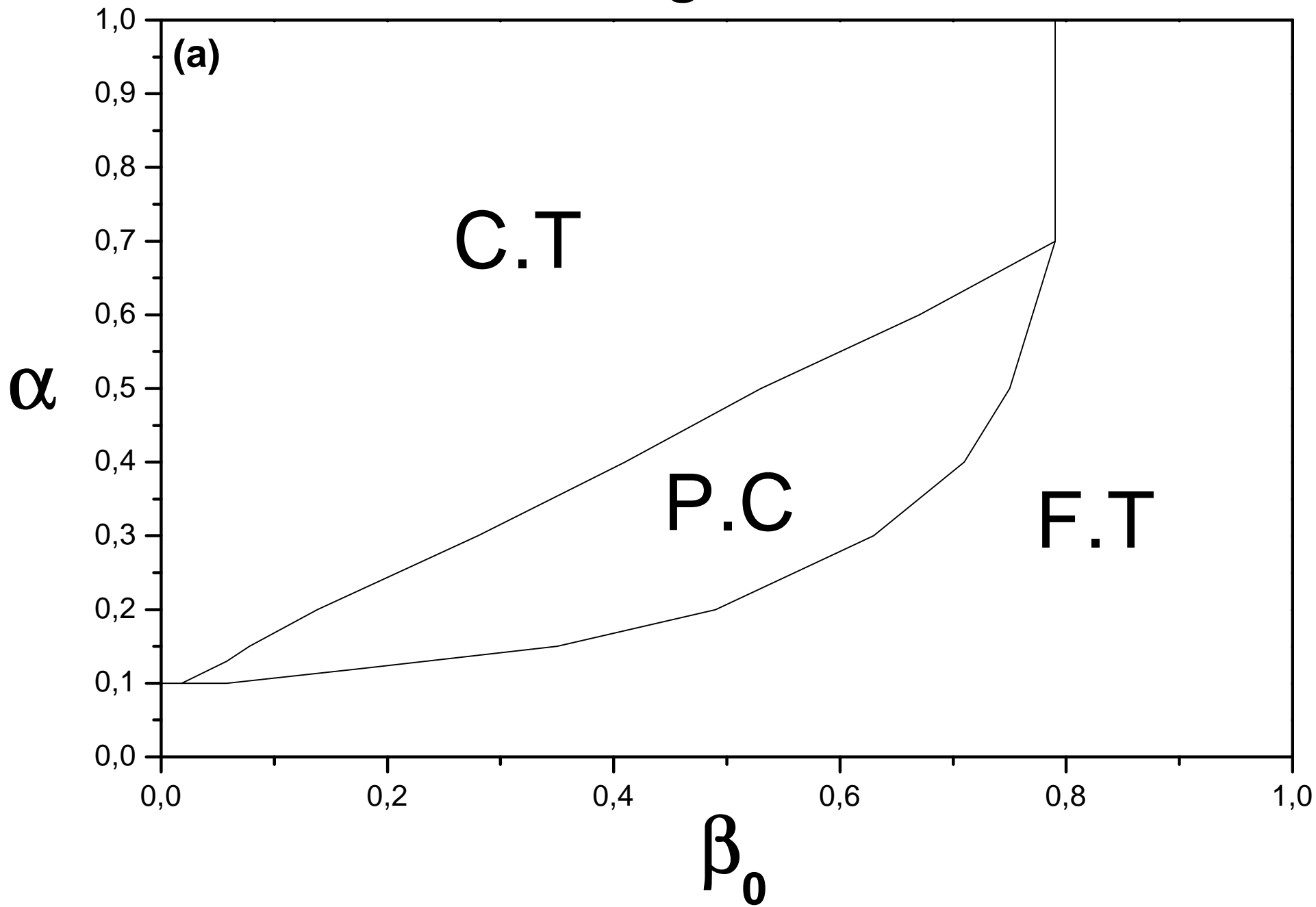
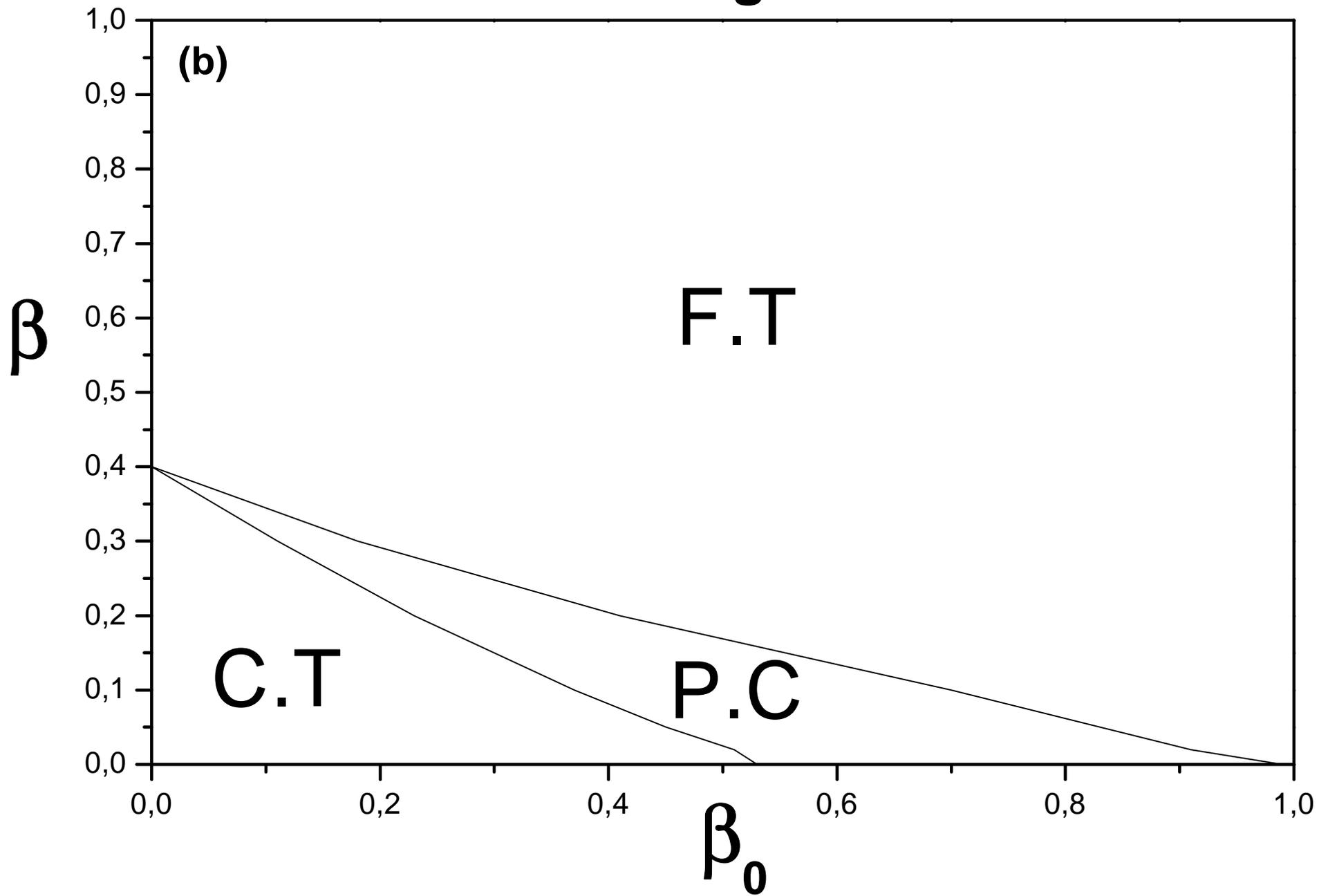
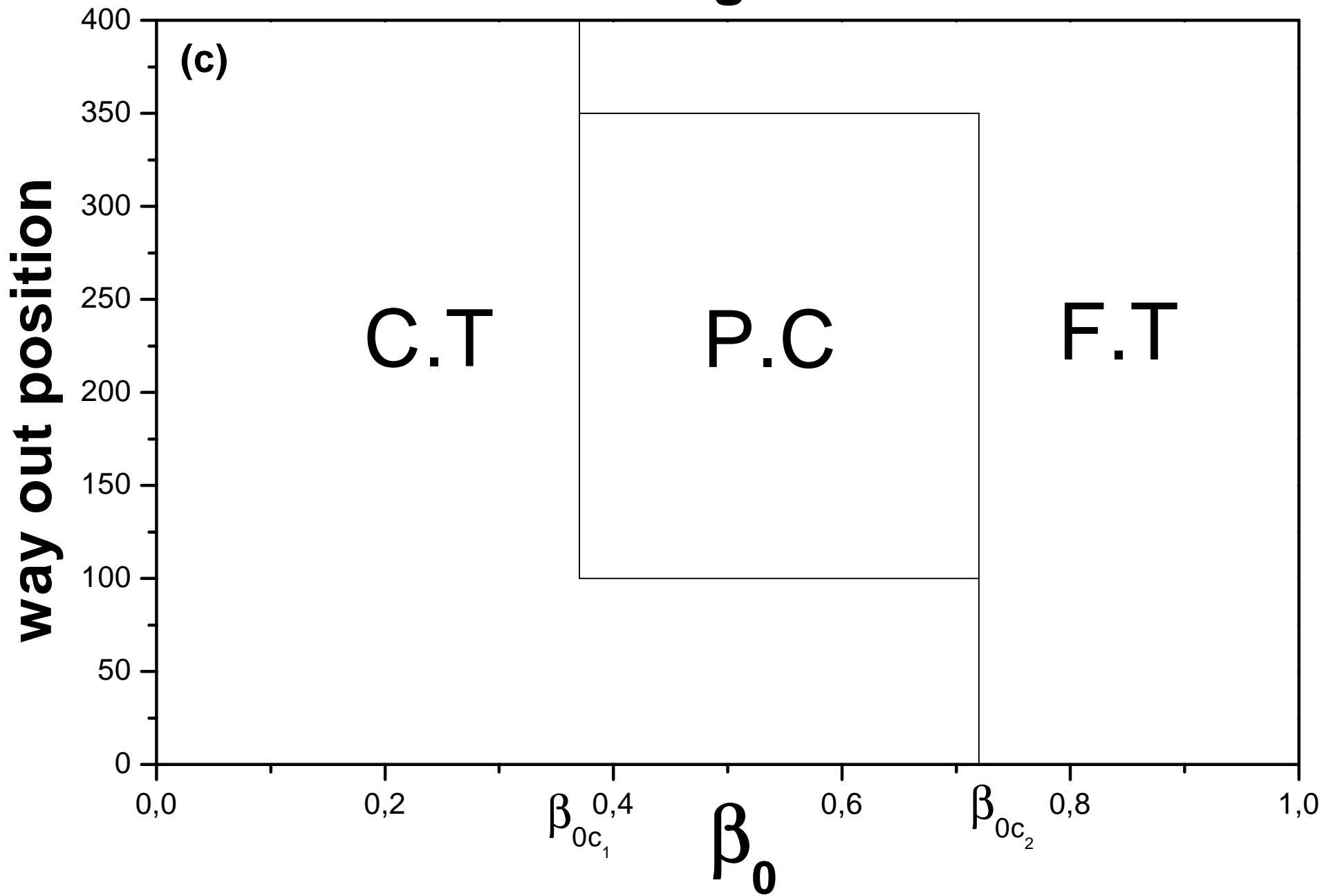
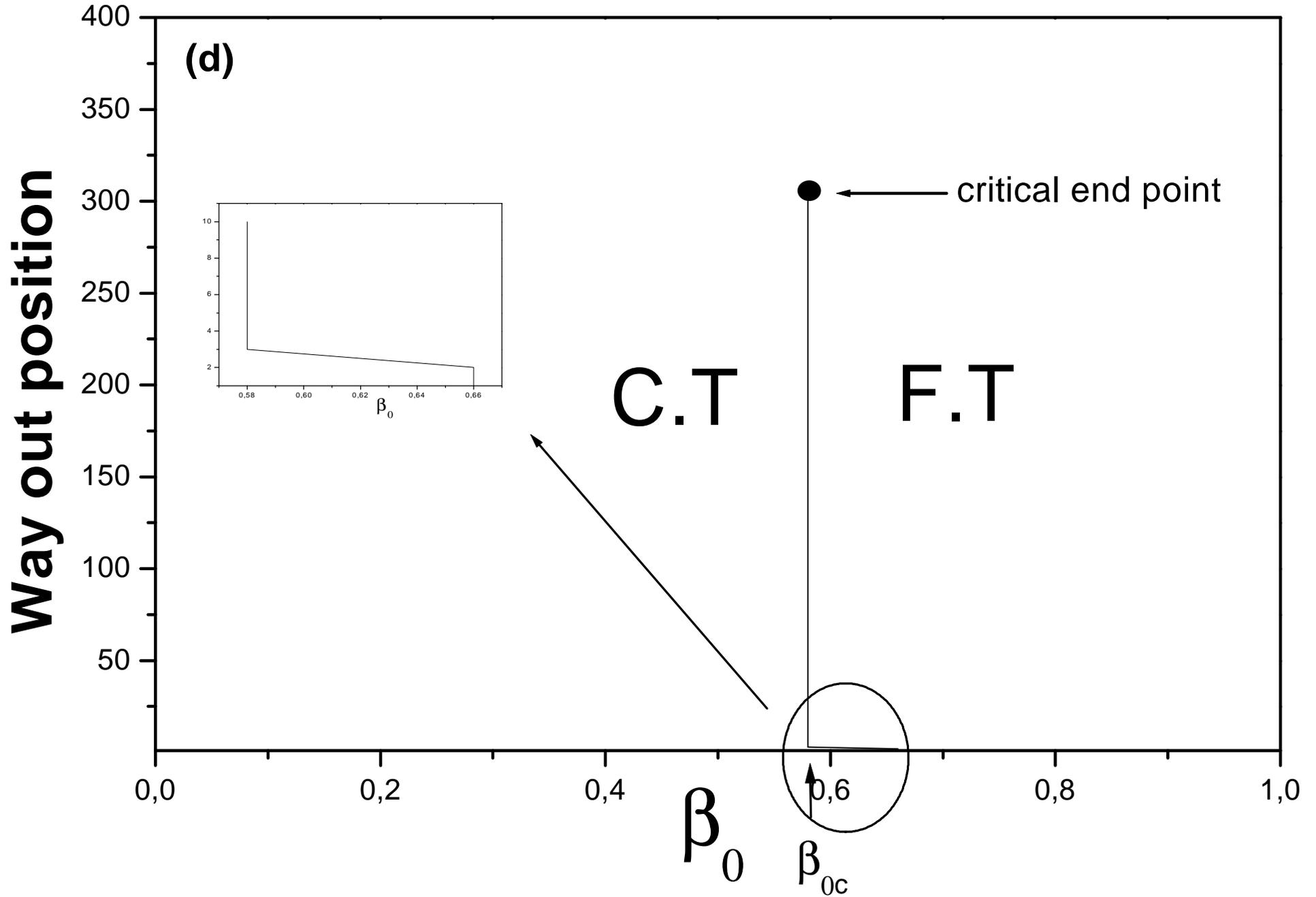
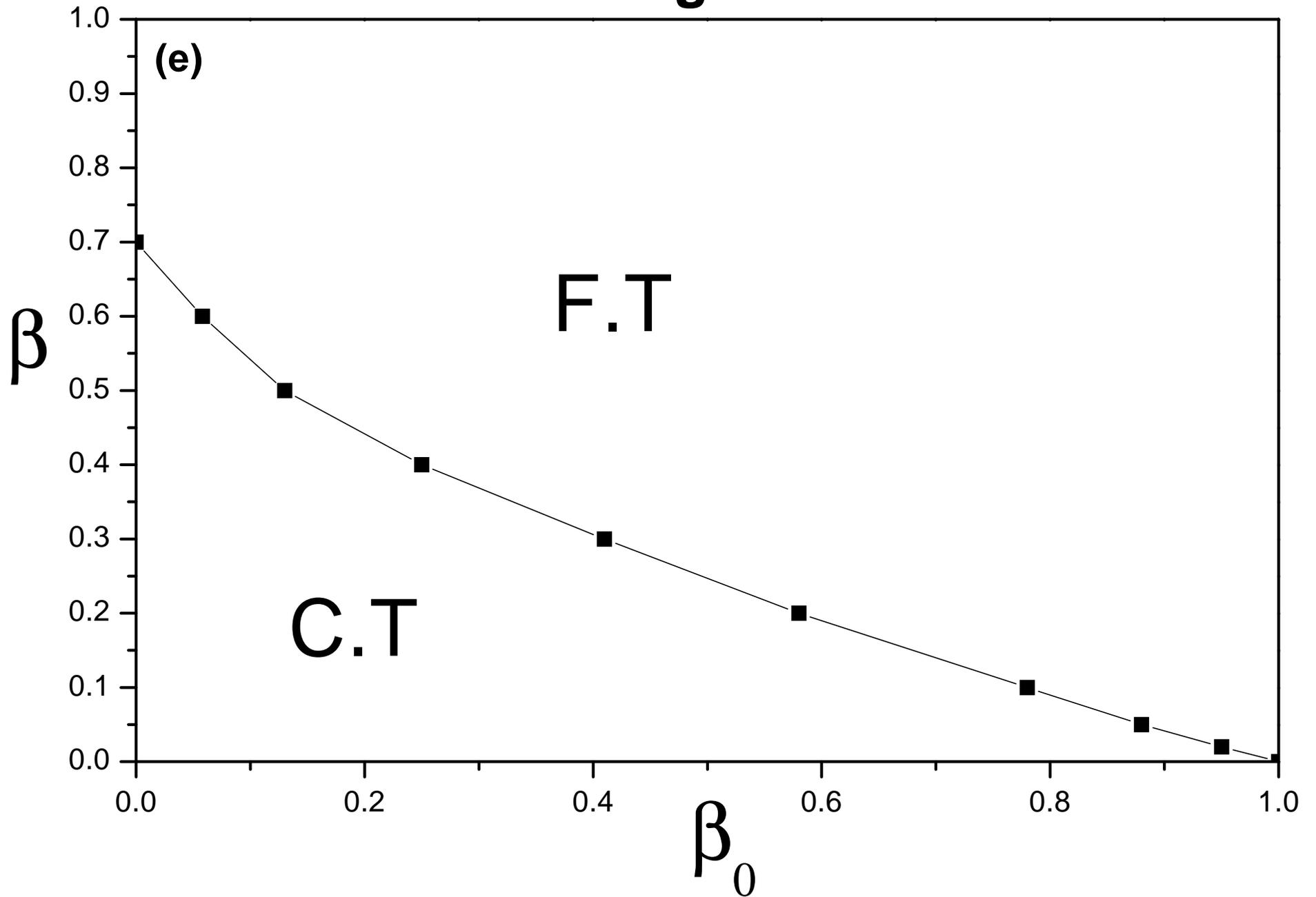
Fig.3 : Phase diagrams (a) $(\alpha,\beta_{0})$ plan for $\beta = 0.1$ and $i_{1}  = 200 $, (b)-(d) present phase diagrams in the $(\beta,\beta_{0})$ plan for $i_{1} = 200$, $\alpha = 0.4$ and $\alpha= 0.7$ respectively, (c)-(e) phase diagrams in the $(i_{1} ,\beta_{0})$ plan for $\alpha = 0.4$, $\beta=0.1$,and  $\alpha = 0.7$, $\beta=0.2$ respectively, for $L = 400$

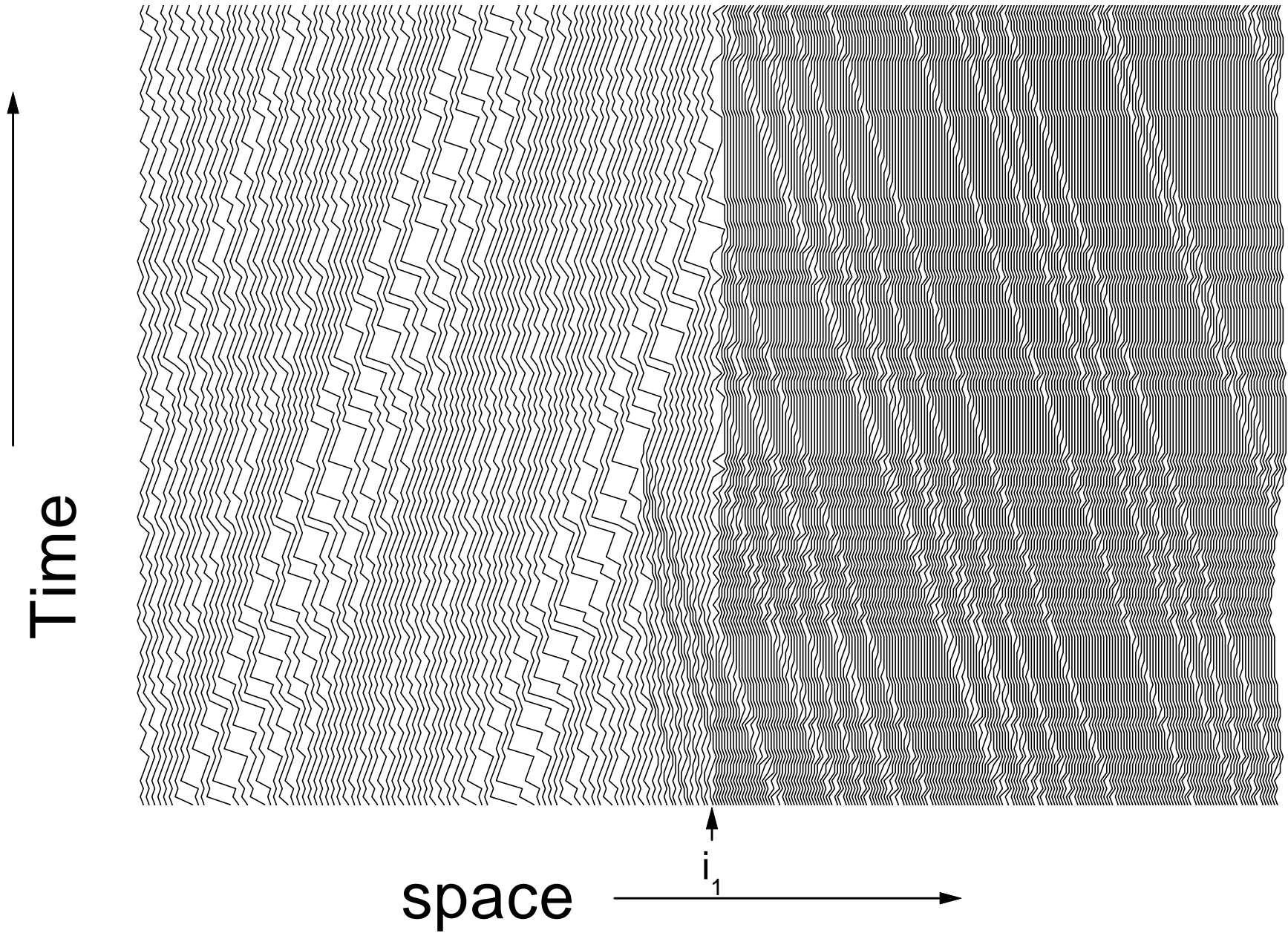
Fig.4: space-Time evolution for $alpha = 0.4$, $beta = 0.1$, $\beta_{0} = 0.6$ and $L = 400$, in case of one way out.

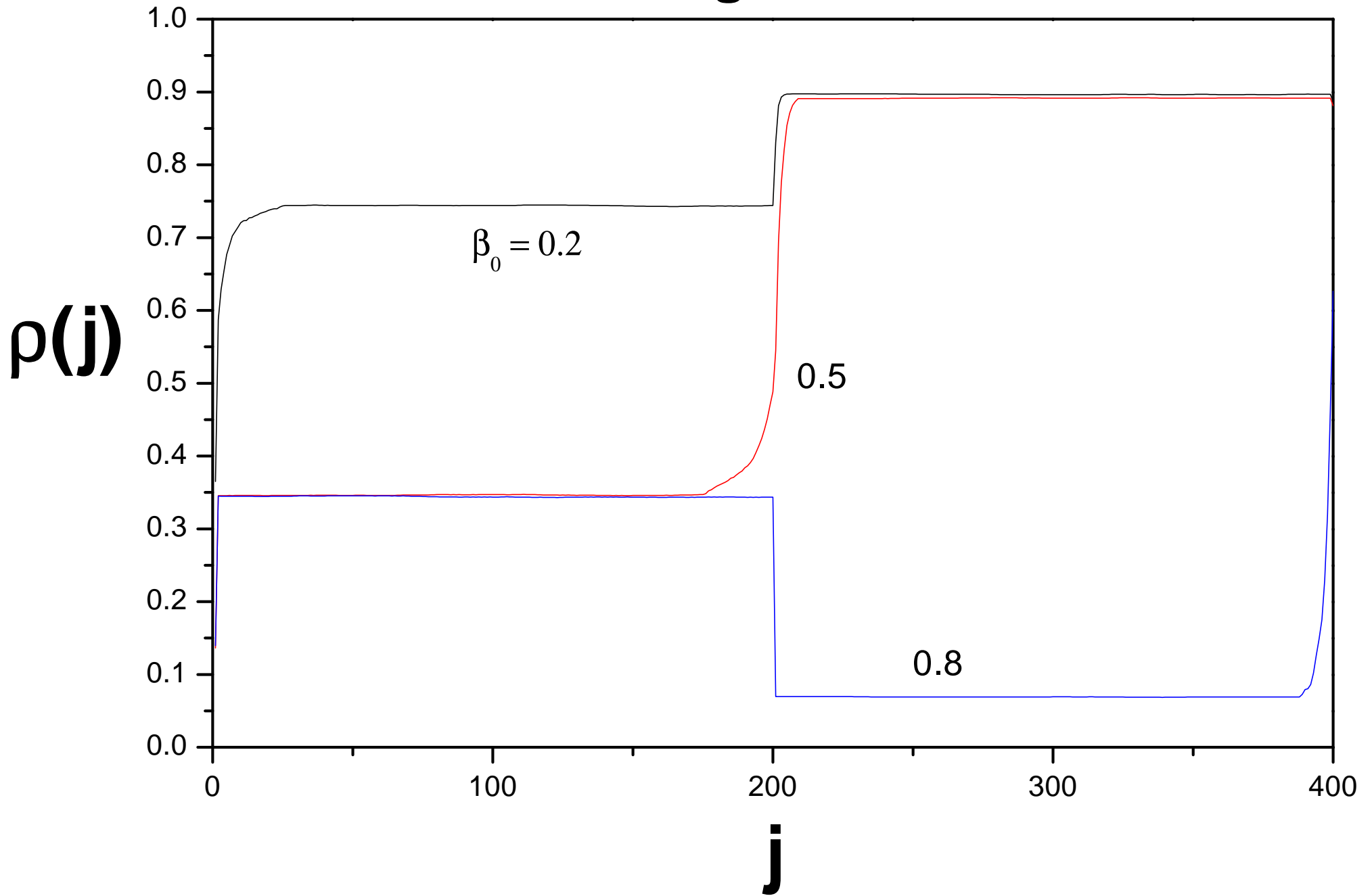
Fig.5: density profile in the case of one way out for $\alpha = 0.4$, $\beta = 0.1$ and for several values of $\beta_{0}$

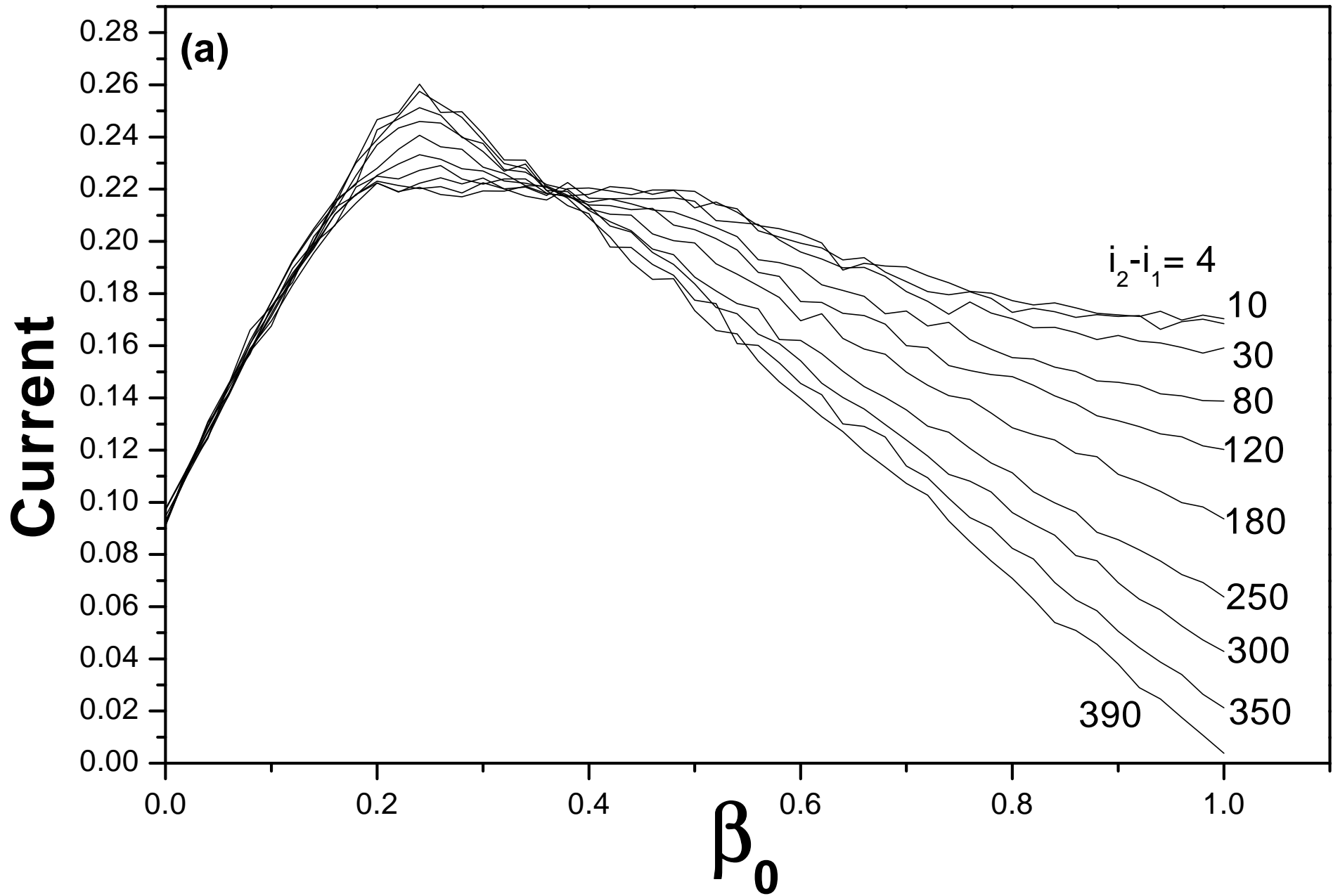
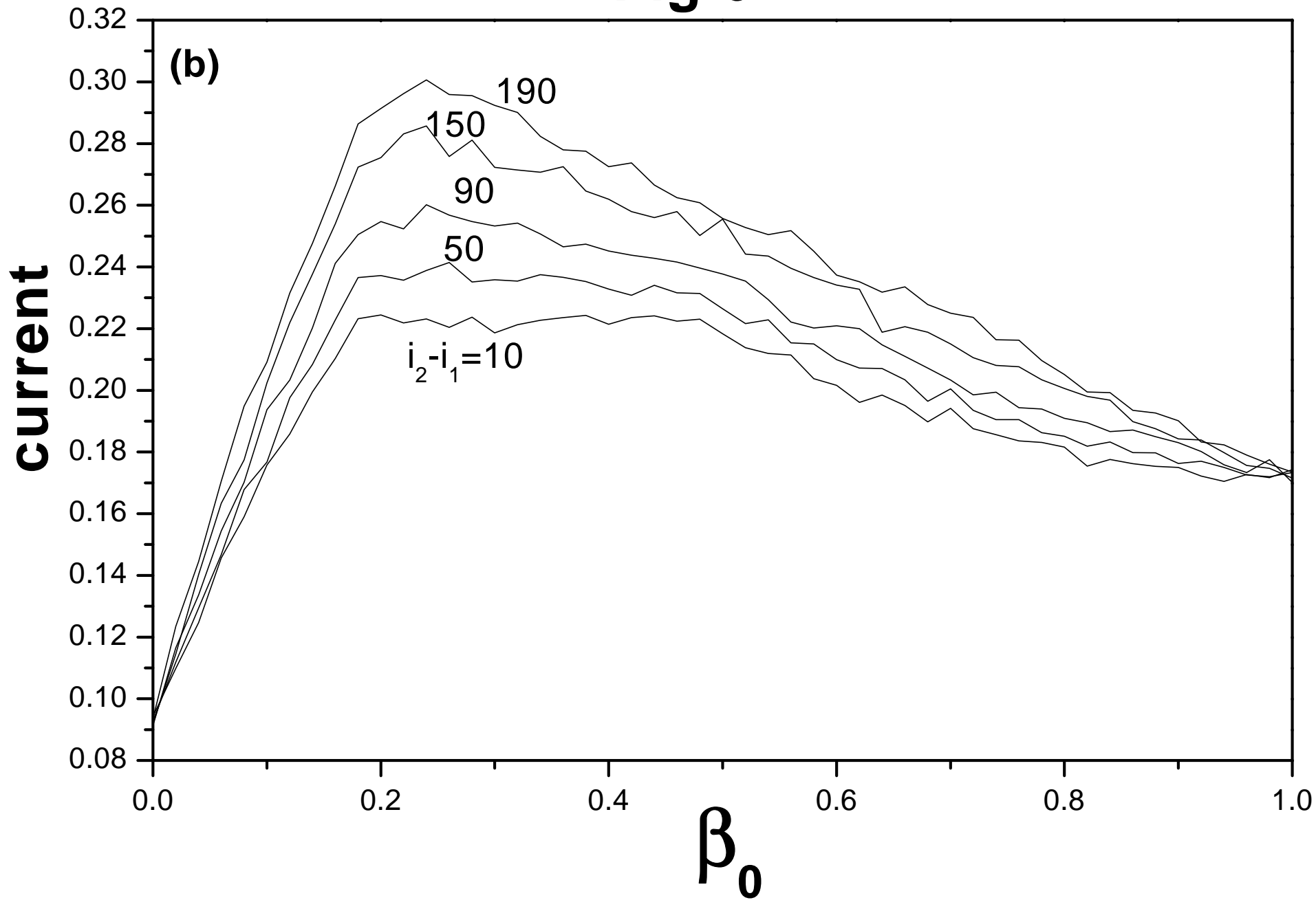
Fig.6 : The variation of the current versus $\beta_{0}$ in the case of two ways out for different values of the distance $i_{2}-i_{1}$ between the two ways, with $\alpha = 0.4$, $\beta = 0.1$ and $L = 400$, (a) the position of the two ways out change symmetrically  with regard to the center of the chain, (b) the first way out is fixed in the middle site and the second change the position
 
\end{document}